\documentclass[11pt]{article}
\newcommand{\be}{\begin{equation}}
\newcommand{\ee}{\end{equation}}
\newcommand{\ba}{\begin{array}}
\newcommand{\ea}{\end{array}}
\newcommand{\bea}{\begin{eqnarray*}}
\newcommand{\eea}{\end{eqnarray*}}
\newcommand{\bean}{\begin{eqnarray}}
\newcommand{\eean}{\end{eqnarray}}
\newtheorem{theorem}{Theorem}[section]
   \newcommand{\bth}{\begin{theorem}}
   \newcommand{\eth}{\end{theorem}}
\newtheorem{lemma}{Lemma}[section]
   \newcommand{\blem}{\begin{lemma}}
   \newcommand{\elem}{\end{lemma}}
\newtheorem{proposition}{Proposition}[section]
      \newcommand{\bprop}{\begin{proposition}}
       \newcommand{\eprop}{\end{proposition}}
\newtheorem{remark}{Remark}[section]
      \newcommand{\brem}{\begin{remark}}
       \newcommand{\erem}{\end{remark}}

\newcommand{\square}{\hfill$\Box$}

\def\ds{\displaystyle}

\def\RR{{\rm I~\hspace{-1.15ex}R} }
\def\CC{\rm \hbox{C\kern-.56em\raise.4ex
         \hbox{$\scriptscriptstyle |$}\kern+0.5 em }}
\def\Box{\leavevmode\vbox{\hrule
     \hbox{\vrule\kern5pt\vbox{\kern5pt}%
           \vrule}\hrule}}

\def\n{\nabla}
\def\eps{\epsilon}
\def\p{\partial}
\def\O{\Omega}
\def\l{\lambda}
\newtheorem{lem}{Lemma}
\newtheorem{prop}{Proposition}
\newtheorem{definition}{Definition}
\newtheorem{thm}{Theorem}
\newtheorem{cor}{Corollary}

\usepackage{graphicx}
\input{epsf}

\newcount\clegende \clegende=0
\def\legende#1{\global\advance \clegende by 1
$$ \hbox{\hss\small Figure \the\clegende~: #1 \hss} $$}

\begin{document}
\title{Asymptotic Expansions for Resonances in the
Presence of Small Anisotropic Imperfections}
\author{M. Gozzi\thanks{ Departement of Mathematics, University of Carthage, Bizerte, Tunisia
(Email:gozzi.maroua@live.fr)}
\and A. Khelifi \thanks{
D\'epartement de Math\'ematiques, Facult\'e des Sciences Bizerte, Universit\'e de
Carthage, Tunisia. (Email: abdessatar.khelifi@fsb.rnu.tn)}}

\maketitle\abstract{We provide a rigorous derivation of an asymptotic formula for perturbations in the
resonance values caused by the presence of finite number of anisotropic imperfections of small shapes
with constitutive parameters different from the background conductivity. The asymptotic expansion is carried out with respect to the size of the imperfections. The main feature of the
method is to yield a robust procedure making it possible to recover information about the location, shape, and material properties of the anisotropic imperfections.}\\



\section{Introduction}
Throughout this paper we consider two or three-dimensional bounded domain
$\Omega$, and assume we have a smooth background conductivity with
small (anisotropic) imperfections of bounded conductivity. The
geometry of the imperfections will take the form of $\eps B$ where
$B$ is some bounded smooth domain. Our goal is to find an
asymptotic expansion for  the  resonance values
of such a domain, with the intention of using the expansion as an
aid in identifying the imperfections. That is, we would like to
find a method for determining the locations and/or shape of small
imperfections by taking resonance measurements.\\
Theoretical modeling of the interaction between harmonic waves and anisotropic objects is of
interest for many physical applications, such as waveguide
optics, nondestructive testing, and remote sensing.\\

Here we discuss the resonances of the transmission problem, which has importance in non-destructive testing of
anisotropic materials, in the whole of $\RR^n$, $n=2,3,$
where the conductivity is assumed to take a positive constant
value outside the domain $\Omega$. The transmission eigenvalue problem is a nonlinear and non-selfadjoint eigenvalue problem
that is not covered by the standard theory of eigenvalue problems for elliptic equations. The eigenvalue problem with the
Dirichlet boundary condition (the Neumann problem, in isotropic media, was
treated in \cite{ammmos}), and the
 scattering problem in $\RR^n
\setminus \Omega$ with the Dirichlet or the Neumann boundary
condition are of equal interest. The asymptotic results for the
eigenvalues or the resonances in such cases can be obtained with
only minor modifications of the techniques presented here and in
\cite{ammmos,khelifi1,khelifi2}, while
the rigorous derivations of similar asymptotic formulae for the
scattering problems for the electromagnetic waves (the full
Maxwell's equations) or for the Stokes equations require further work.

In this paper we concentrate on deriving rigorously the asymptotic
expansion for resonances. This work is considerably different from that in \cite{AKhelifi,ammmos,cakoni1,cakoni2,khelifi1,khelifi2} for the eigenvalue problem. In \cite{AKhelifi,khelifi2}, we combined the
expansions derived in \cite{fenmosvog} and an additional lemma
with a theorem by Osborn \cite{osb} about the convergence of
eigenvalues of a sequence of compact operators.\\

The novelty of this paper, is that it leads to analysis of resonance problems in the presence of multiple anisotropic imperfections.
By referring to previous works, the resonance problem is more difficult because the resonance
values are not the eigenvalues of a set of compact operators. They
can, however, be viewed as singular values of a sequence of
meromorphic operator functions. This is achieved by rewriting the
problem in terms of integral equations on the boundary of the
domain. Then, by using complex operator theory, we derive a
formula for the convergence of the resonance values which is in
the spirit of the theorem of Osborn. This then leads to an
asymptotic expansion for the resonances which is similar to that
for the eigenvalues.

The leading order term in the asymptotic expansion contains information
about the location, shape, and material properties of the inhomogeneities.\\

The paper is organized as follows. In Section $2$, we model the search of resonances of the transmission problem in the whole of $\RR^n$ as two linear spectral problems (\ref{reseps})
and (\ref{res}). Then, in Section $3$, we reformulate (\ref{reseps}) and (\ref{res}) as two systems of integral
equations depending on the small parameter $\epsilon$ which is the scale factor associated to anisotropic imperfections. By the analytic Fredholm theory,  we transform these systems into the determination of the poles of two memorphically continued inverse integral operator-valued functions $T^{-1}(w)$ and $T_\epsilon^{-1}(w)$ in the complex plane. Section 4 is devoted to the rigorous derivation of asymptotic expansions of the resonances as $\epsilon$ goes to zero.

\section{Problem formulation}
\begin{figure}[!htb]
\begin{center}
\includegraphics[width=7cm,height=4cm]{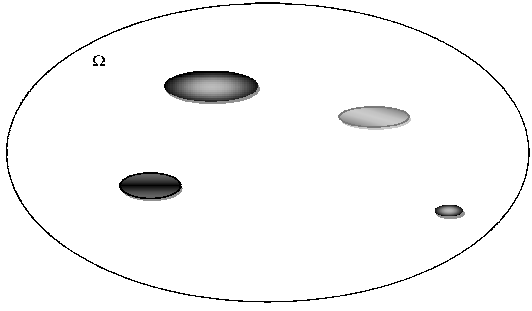}
\end{center}
\caption{Configuration of the domain $\Omega$ with four imperfections (m=4).} \label{fig1}
\end{figure}

In this paper we consider two or three-dimensional bounded domain
$\Omega \subset \RR^n$ ($n=2$ or $3$) with $C^2$-smooth boundary $\partial\Omega$, and let $\nu$ denotes the outward unit normal vector on $\partial\Omega$.
  We suppose that $\Omega$ contains a finite number $m$ of (possibly anisotropic) bounded imperfections (see Figure \ref{fig1}). The geometry of each one is of the form $D_i=z_i +\eps B_i$, with a smooth and simply connected
boundary $\partial D_i$, where $B_i\subset \RR^n$ is a regular enough bounded domain representing the volume of the imperfection, $z_i\in \RR^n$ is the vector position
of its center and $\eps\in \RR_+$ is the scale factor. The total collection
of imperfections thus takes the form $D_\eps:=\cup_{i=1}^mD_i$, with the following assumption:
\begin{equation}\label{zass}
\bar{D}_i\cap \bar{D}_j=\emptyset\quad \forall i\neq j,\quad \mbox{and } 0<d_0\leq\hbox{dist}(z_j,\partial \Omega),
\end{equation}
where $d_0$ is a positive constant.

We assume that the constitutive parameters of the media in $D_\eps$ are represented by a real-valued symmetric
matrix $\gamma_{D_\eps}=(\gamma_D^{ij})\in C^1 (D_\eps,\RR^{n\times n})$ such that $\xi\cdot  \gamma_{D_\eps}(x)\xi\geq a_{0}|\xi|^2>0$ for almost all $x \in D_\eps$
 and all $\xi\in{\bf C}^n$, and where $a_0$ is a positive constant. Outside $D_\eps$ the
background media is isotropic. We denote by $A_\eps$ the constitutive parameters of the anisotropic background $\RR^n$ given by
\be\label{gamma_eps0}\gamma_\eps(x): = \left\{
                  \begin{array}{lll} \gamma_{D_\eps} (x)& x\in D_\eps\\
\gamma_1I_n &
x\in \O\setminus \bar{D}_\eps \\ \gamma_2 I_n& x\in \RR^n\setminus \bar{\Omega}
  \end{array}
                \right. \ee
where $I_n$ is the identity matrix in $\RR^n$, and $\gamma_1$, $\gamma_2$ are two positive constants. For any regular function $v$ and in terms
of $\gamma_\eps$, we have $\displaystyle\nabla\cdot \gamma_\eps\nabla v:=\sum_{i,j=1}^n\frac{\partial }{\partial x_i}\big(\gamma_\eps^{ij}\frac{\partial v}{\partial x_j}\big).$\\

We are interested in calculating resonances or scattering frequencies $\l_\eps$ of the
following problem in dimension $n=2$ or $3$:
\begin{equation}\label{reseps} \n\cdot\gamma_\eps\n u_\eps +\l^2_\eps u_\eps=0
\ \ \ \ \ \mbox{in}\ \ \  \RR^n
\end{equation}
with $u_\eps\in H_{loc}^{1}(\RR^n)$ satisfying the following radiation condition at infinity: $$ |\p_r u_\eps -i
\frac{ \l_\eps}{\sqrt{\gamma_2}} u_\eps |\leq {c\over{r^{n-1}}}.$$
where $r=|x|$, and the coefficient $\gamma_\eps$ takes on one value outside
some bounded domain $\O$, another inside $\O$, and a different value
inside the small (anisotropic) imperfections $D_i$ in the interior of $\O$.\\ 

Our goal is to express $\l_\eps$ in an asymptotic expansion in terms of
the resonances of the unperturbed problem:
\begin{equation}\label{res} \n\cdot\gamma(x)\n u +\l^2 u=0
\ \ \ \ \ \mbox{in}\ \ \  \RR^n
\end{equation}
$$ |\p_r u -i \frac{\l}{\sqrt{\gamma_2}} u |\leq
{c\over{r^{n-1}}}.$$ where
$$\gamma(x):=\gamma_1\chi(\O)I_n+\gamma_2\chi(\RR^n\setminus \bar{\O})I_n,\quad (\mbox{for }\epsilon=0),$$

where $\chi(\O)$ (resp. $\chi(\RR^n\setminus \bar{\O})$) is the characteristic function of $\O$ (resp. of $\RR^n\setminus \bar{\O}$), and $\gamma_1$ and $\gamma_2$ are given by (\ref{gamma_eps0}).\\

To the best of our knowledge, this work is the first one to solve the resonance problems, in the presence of multiple (with finite number) anisotropic imperfections, but by referring to other existing related problems.\\

Now, we can remark that
\begin{remark}\label{rem1}\begin{enumerate}
             \item For each $j=1,\cdots,m$, we put $\gamma_{D_\eps}(x)=\gamma_{D_j}(x)$ for all $x\in D_i$ and where $\gamma_{D_j}=(\gamma_{D_j}^{il})_{1\leq i,l\leq n}$ is a symmetric
matrix representing the constitutive parameters of the media in $D_j$ are represented by a real-valued symmetric
matrix $\gamma_{D_j}=(\gamma_{D_j}^{il})\in C^1 (D_j,\RR^{n\times n})$ such that $\xi\cdot  \gamma_{D_j}(x)\xi\geq a_{j}|\xi|^2>0$ for almost all $x \in D_j$ and all $\xi\in{\bf C}^n$, and where $a_j$ is a positive constant such that $\displaystyle a_0=\min_{1\leq j\leq m}a_j$; where $a_0$ was given before.
\item For a function $u\in C^1(D_\eps)$ we define the conormal derivative by $$\displaystyle \frac{\partial u}{\partial\nu_{\gamma_\epsilon}}(x) :=	\nu(x)\cdot \gamma_\eps(x)\nabla u(x),\quad x\in\partial D_\eps.$$
           \end{enumerate}
\end{remark}

In terms of the previous notations, we have the following definition.
\begin{definition}\label{def1}
Regarding (\ref{reseps}), we may define:
\begin{itemize}
  \item [1)] The values of $\lambda_\eps\in \CC$ for which the following resonance problem
$$\left\{
  \begin{array}{ll}
    \nabla\cdot\gamma_\eps\nabla u_\eps+\lambda_\eps^2 u_\eps=0 & \hbox{in }\Omega \\
    u_\eps=0 & \hbox{on }\partial\Omega
  \end{array}
\right.$$
 has a nontrivial solution $u_\eps\in H^2(\Omega)\cap H_0^1(\Omega)$ are called the resonant frequencies.
\item [2)]The values of $\lambda_\eps\in \CC$ for which the following resonance problem
$$\left\{
  \begin{array}{ll}
    \nabla\cdot\gamma_\eps\nabla u_\eps+\lambda_\eps^2 u_\eps=0, & \hbox{in }\RR^n\backslash\bar{\Omega} \\
    u_\eps=0, & \hbox{on }\partial\Omega\\
|\p_r u_\eps -i
\frac{ \l_\eps}{\sqrt{\gamma_2}} u_\eps |\leq {C\over{r^{n-1}}}.&
  \end{array}
\right.$$
 has a nontrivial solution $u_\eps$ are called the scattering frequencies.
\end{itemize}
\end{definition}
The interior transmission eigenvalue problem (ITEP)
corresponding to the scattering problem for an anisotropic inhomogeneous medium
in $\RR^n,$ $n = 2, 3$ reads:
\begin{equation}\label{ITEP}\left\{
  \begin{array}{ll}
    \nabla\cdot\gamma_1\nabla u+\lambda^2 u=0 & \hbox{in }\Omega\backslash\overline{D}_\eps \\
 \nabla\cdot \gamma_{D_\eps}\nabla u_\eps+\lambda_\eps^2 u_\eps=0 & \hbox{in }D_\eps\\
u_\eps-u=h_1 & \hbox{on }\partial D_\eps\\
\frac{\partial u_\eps}{\partial \nu_{\gamma_\eps}}-\frac{\partial u}{\partial \nu}=h_2 & \hbox{on }\partial D_\eps\\
u=0 & \hbox{on }\partial\Omega,
  \end{array}
\right.\end{equation}
with the given data $h_1\in¸H^{1/2}(\partial D_\eps)$ and $h_2\in H^{-1/2}(\partial D_\eps)$. The interior Neumann problem (on $\partial\Omega$) can be treated in essentially the same way as the interior Dirichlet problem.\\

Since $\gamma_{D_{\eps}}$ is symmetric, positive definite matrix, it follows that, in a well determined basis in $\RR^n$, we can write $\gamma_{D_{\eps}}(x):=diag\{k_j(x),\quad j=1,\cdots,n\}$; where $k_j(x)$ ($x\in D_\eps$) is a positive valued-function. Consequently, by using Remark \ref{rem1} we can write $\gamma_{D_{i}}(x):=diag\{k_j^i(x),\quad j=1,\cdots,n\}$, for $i=1,\cdots,m$, where $k_j^i(x)$ ($x\in D_i$) is a positive valued-function.\\
On the other hand, by assuming that $k_j^i$ is constant on $D_i$, we may refer to the existing works (but for isotropic inclusions) \cite{AKhelifi,ammmos,volvol} to define the polarization tensor,
$M^{(i)}$, which is an $n\times n$, symmetric, positive definite matrix
given by
\begin{equation}
M^{(i)}_{jl}=|B_i|\delta_{jl}+ \left({\gamma(z_i)\over{tr(\gamma_{D_{i}})}}-1\right)
\int_{\partial B_i}y_j{\partial\phi_l^+\over{
\partial\nu}}d\sigma_y \label{Meq}
\end{equation}
where $\displaystyle tr(\gamma_{D_{i}}):=\sum_{j=1}^nk_j^i$ means the trace of the matrix $\gamma_{D_{i}}$, $\gamma(z_i)$
is the background conductivity and the point $z_i$, and for $l=1,\ldots m$;
$\phi_l(y)$ is the unique function which satisfies
\begin{eqnarray} \Delta\phi_l &=& 0\ \ \ \ \mbox{in}\ \ \ B_i\label{phi1} \\
\Delta\phi_l &=& 0\ \ \ \ \mbox{in}\ \ \ \RR^n\setminus B_i\nonumber \\
\gamma(z_i)
{\partial \phi_l^+\over{\partial\nu}}- tr(\gamma_{D_{i}}){\partial\phi_l^-\over{\partial\nu}}
&=& - tr(\gamma_{D_{i}})\nu_l \ \ \ \ \mbox{on}\ \ \ \partial B_i\nonumber
\end{eqnarray}
with $\phi_l$ continuous across $\partial B_i$ and $\lim_{|y|\rightarrow\infty} \phi_l(y)=0.$ Here $\nu=(\nu_1,\cdots,\nu_n)$ still denotes the outward unit
normal to $\partial B_i$; superscript $+$ and $-$ indicate the limiting values as we approach $\partial B_i$ from outside $B_i$, and from inside $B_i$, respectively. Note that the polarization tensor depends on the conductivity,
size, and shape of the imperfection.\\

It is  known that the set of resonances $\{ \l\}$ of the
transmission problem (\ref{res}) is discrete and symmetric about the imaginary
axis if $\gamma_1$ and $\gamma_2$ are real. Further, it can be
easily seen that all the resonances $\{ \l \}$ are in the lower
half-space $Im \l < 0$. They can be found explicitly for a
sphere and a constant conductivity $\gamma$ and are connected in
this case with the zeros of certain Bessel functions. More
elaborate results assert that for a strictly convex domain
$\Omega$ in $\RR^3$ and a constant conductivity $\gamma_1 >
\gamma_2$ the resonances $\{ \l\}$ accumulate on the real axis as
$|\l| \rightarrow + \infty$.  One would expect that for such a
situation the resonances (at least for a sequence) come rapidly to
the real axis as $|\l| \rightarrow + \infty$. This is in fact
the case and was shown in a recent work by Popov and Vodev
\cite{popvod1}, \cite{popvod4}, \cite{popvod2}. For a strictly
convex domain it has been proved by Cardoso, Popov and Vodev
\cite{carpopvod} that if $\gamma_1 < \gamma_2$ there are no
resonances in  a strip above the real axis. We know of three
methods in the literature for finding the resonances of the
transmission problem. One is the method of the Rayleigh-Ritz type,
well known in quantum chemistry \cite{rei}. Another closely
related method consists of approximating the exact radiation
condition by an appropriate boundary condition (Dirichlet
condition or something like the on-surface boundary conditions
\cite{kri}, \cite{jones}) on a large sphere and then computing the
eigenvalues of the resulting problem \cite{docho}. The strategy in
the third method is to solve the time-dependent wave equation for
any appropriate initial data at $t=0$. The resonances are then
calculated from the asymptotic expansion of the solution for fixed
point $x$ as $t \rightarrow + \infty$ \cite{weimadstr}. Little is
known about a constructive method for finding the resonances of
the transmission problem like the variational principle for
eigenvalues of the interior problems. For material on resonances
of the scattering problem with Dirichlet or Neumann boundary
condition on the boundary of the obstacle $\Omega$ we refer the
reader for example to \cite{cakoni1}, \cite{cakoni2}, \cite{laxphi}, \cite{tay}, \cite{sjo} and the references
therein. For other important investigations concerning the interplay between resonances and geometry, we refer the readers to  well-know works of Broer et al. \cite{Broer1,Broer2}, and that of Zworski \cite{zwo}.

\section{Integral equations and operator convergence}

In this section will reformulate the problems (\ref{res}) and (\ref{reseps})
in terms of integral equation operators and show operator convergence. A similar integral equation method may be applied to solve the eigenvalue problems given by Definition \ref{def1} or the interior transmission eigenvalue problem (ITEP) defined in (\ref{ITEP}).
The use of integral equations is a convenient tool for a number of
investigations in scattering theory \cite{colton}, \cite{tay}, \cite{poi},
\cite{popvod1}, \cite{popvod2}, \cite{popvod3}, and for basic approaches by means of operators we may refer to \cite{Kaashoek}. We use it here to characterize the resonances as poles of a
meromorphic operator-valued function on the whole complex plane.
This characterization is the key point in deriving our asymptotic
formulae.
\subsection{Boundary integral formulation}
Assume that $\gamma(x):=\gamma_1\chi(\O)I_n+\gamma_2\chi(\RR^n\setminus \bar{\O})I_n,$ (for $\epsilon=0$). It is well know that for $\omega$ not a resonance value for
 $$(\gamma_2 \Delta + \omega^2 ),$$
there exists a unique Green's function $G^\omega (x,y)$ satisfying
\begin{equation} (\gamma_2\Delta_y +\omega^2 )G^\omega (x,y)=-\delta_x(y)
\label{Geq} \end{equation} along with the radiation condition $$
|\p_{r} G^\omega -i\frac{\omega}{\sqrt{\gamma_2}} G^\omega |\leq
{C\over{r^{n-1}}}.$$ We actually know $G^\omega$ explicitly, when
$n=3$ $$G^\omega (x,y)
={e^{i{\omega\over{\sqrt{\gamma_2}}}|x-y|}\over{ 4\pi \gamma_2
|x-y| }}$$ and $$G^\omega (x,y) = \frac{1}{\gamma_2}
H^{(1)}_0(\frac{\omega}{\sqrt{\gamma_2}} |x -y|)$$ for $n=2$.
Here, $H^{(1)}_0$ is the Hankel function of order $0$.  By
integration by parts, we see that if the pair ( $u(x)$, $\l$) is a
nontrivial solution to (\ref{res}), outside of $\O$ $u(x)$ can be
expressed as
\begin{equation} u(x)= \int_{\p\O}\gamma_2 \p_{\nu_y} G^\l (x,y) u(y)
d\sigma_y - \int_{\p\O}\gamma_2G^\l (x,y) \p_{\nu} u(y)|_{\p\O^+}d\sigma_y,
\label{int1}\end{equation} where the second term is continuous up
to the boundary but the first is not. Define the operator
$N^\omega$ to be the interior Dirichlet to Neumann map, i.e.
$$N^\omega: H^{1/2}(\p\O)\rightarrow H^{-1/2}(\p\O).$$ where
$$N^\omega (f):= \p_{\nu} v|_{\p\O^-}$$ for $v$ the solution to
$$\gamma_1\Delta v+ \omega^2 v =0 \ \ \ \ \mbox{in}\ \ \O $$
$$v=f\ \ \ \ \mbox{on}\ \ \ \p\O.$$ The above boundary value
problem is well-posed in $H^1(\Omega)$ for any $\omega$ not a Dirichlet
eigenvalue of the operator $\gamma_1 \Delta$ in the bounded domain
$\Omega$.
Hence the operator $N^\omega$ is then well-defined on $\CC$ outside the
set of these (real) eigenvalues.

Now, from the transmission condition $$\gamma_1\p_{\nu}
u|_{\p\O^-}=\gamma_2\p_{\nu} u|_{\p\O^+},$$ we get, by taking the
limit of (\ref{int1}) as $x\rightarrow \p\O^+$ (see Taylor
\cite{tay}  or Folland \cite{fol}),
$$(1-{\gamma_2\over{2}})u|_{\p\O}=\gamma_2\int_{\p\O}\p_{\nu} G^\l
u|_{\p\O} - \gamma_1\int_{\p\O}G^\l N^\l (u|_{\p\O}).$$ If the
single and double layer potential operators
 $$S^\omega: H^{-1/2}(\p\O)\rightarrow H^{1/2}(\p\O)$$ and
 $$D^{\omega}: H^{1/2}(\p\O)\rightarrow H^{1/2}(\p\O)$$
are defined by
 $$S^\omega: g\mapsto \int_{\p\O} G^\omega(x,y) g(y)d\sigma_y$$
and
 $$D^\omega: f\mapsto \int_{\p\O}\p_{\nu_y} G^\omega(x,y) f(y)d\sigma_y ,$$
we see that $u(x)|_{\p\O}$ satisfies
$$\left( (1-{\gamma_2\over{2}})I-\gamma_2 D^\l +\gamma_1 S^\l N^\l
\right)(u|_{\p\O})=0$$
where $I$ is the identity operator. This suggests we define the
operator $T$ with complex parameter $\omega$ ,
$$T(\omega): H^{1/2}(\p\O)\rightarrow H^{1/2}(\p\O)$$
by
\begin{equation}
T(\omega)= (1-{\gamma_2\over{2}})I-\gamma_2 D^\omega +\gamma_1
S^\omega N^\omega. \label{T} \end{equation} Then clearly the
resonances of (\ref{res}) are values of $\lambda$ for which $T$
has no inverse, since there exists $u\in H^{1/2}(\p\O)$ such that
$T(\l)u=0.$

Let $\langle., .\rangle$ be the $L^2(\partial \Omega)$-inner
product. By Green's formula it is easy to see that for any $u,
v \in H^{1/2}(\partial \Omega)$ $$ \langle N^\omega u, v \rangle =
\langle u, (N^\omega)^* v\rangle$$ where the $L^2$-adjoint operator of
$N^\omega$ is given by  $$(N^\omega)^* = N^{\overline{\omega}}.$$
Moreover, one may check that we have
$$(S^\omega)^*:  H^{-1/2}(\p\O)\rightarrow H^{1/2}(\p\O)$$
given by
$$(S^\omega)^* = S^{- \overline{\omega}},$$ and
$$(D^\omega)^*:  H^{-1/2}(\p\O)\rightarrow H^{-1/2}(\p\O)$$
given by
$$(D^\omega)^*g = \int_{\p\O}\p_{\nu_x}G^{-\overline{\omega}}(x,y)
g(y)d\sigma_y.$$
Hence the dual of $T$,
$$T^*(\omega):H^{-1/2}(\p\O)\rightarrow H^{-1/2}(\p\O)$$
is equal to
$$T^*(\omega)=
(1-{\gamma_2\over{2}})I -\gamma_2 (D^\omega)^* +\gamma_1
N^{\overline{\omega}}S^{-\overline{\omega}}.$$
Furthermore, from elliptic regularity it
follows that $$N^\omega - N^0 : H^{1/2}(\partial \Omega)
\rightarrow H^{- 1/2}(\partial \Omega)$$ is compact and so,
$$T(\omega) - T(0) : H^{1/2}(\partial \Omega) \rightarrow
H^{1/2}(\partial \Omega)$$ is also a compact operator. By the
analytic Fredholm theory \cite{pros}, \cite{tay}  the resonances
of (\ref{res}) are poles of the memorphically continued inverse
$T^{-1}(\omega)$ to the whole complex plane if $n = 3$. This is
also valid for $n=2$, but then we have to continue
$T^{-1}(\omega)$ to the Riemann surface for $\log \omega$ rather
than $\CC$.

Similarly, if the
pair ($u_\eps,\l_\eps$) is a nontrivial solution of
(\ref{reseps}), we can express $u_\eps$ in the exterior of $\O$ by
$$u_\eps (x)=\gamma_2\int_{\p\O} \p_{\nu} G^{\l_\eps} u_\eps|_{\p\O}-
\gamma_2\int_{\p\O} G^{\l_\eps}\p_{\nu} u_\eps|_{\p\O^+}.$$ Define now
the interior Dirichlet to Neumann map corresponding to the medium
with an inhomogeneity, $$N^\omega_\eps :H^{1/2}(\p\O)\rightarrow
H^{-1/2}(\p\O)$$ $$N^\omega_\eps (f):=\p_{\nu} v_\eps|_{\p\O^-}$$ where $v_\eps$
is the solution to $$\n\cdot\gamma_\eps\n v_\eps+\omega^2 v_\eps=0
\ \ \ \ \mbox{in}\ \ \ \O$$ $$v_\eps=f\ \ \ \ \mbox{on}\ \ \
\p\O.$$ again, by taking the limit as $x\rightarrow \p\O^-$,  we
see that $u_\eps|_{\p\O}$ is a nontrivial solution to $$\left(
(1-{\gamma_2\over{2}})I-\gamma_2 D^{\l_\eps}+\gamma_1 S^{\l_\eps}
N^{\l_\eps}_{\eps}\right)(u_\eps|_{\p\O})=0.$$ Now it seems
natural to define
\begin{equation} T_\eps(\omega)=  (1-{\gamma_2\over{2}})
I-\gamma_2 D^\omega+\gamma_1 S^{\omega}
N^\omega_\eps ,\label{Teps} \end{equation}
so that the resonances of (\ref{reseps}) are  poles of
 the complex meromorphic $T^{-1}_\eps(\omega)$.
We have thus reduced the analysis
of the perturbed and unperturbed scattering problems to the asymptotic
analysis of the poles of $T^{-1}_\eps(\omega)$ in a neighborhood of the
poles of $T^{-1}(\omega)$.

\subsection{Operator convergence}
We will need to obtain pointwise convergence estimates for $T_\eps$ to
$T$ and a convenable asymptotic expansion in
terms of $\epsilon$. These operators are (constant permitivity)
examples of those
described in \cite{ammlakmos} with the exception that in that paper the
operators correspond to the special case $\gamma_2=1$. We also mention
that the authors did not emphasize the dependence of the operators on $\omega$
since the frequency was assumed to be fixed throughout that work.
Since
$$T_\eps(\omega)-T(\omega)=\gamma_1 S^\omega (N_\eps^\omega
-N^\omega),$$
the difference of the operators here is just a constant multiple of
the difference of the operators in Proposition 1 of \cite{ammlakmos}. We may use the following definition.
\begin{definition}(Collectively compact) A set of operators, $\{K_p;\quad p\in I\}$, is collectively compact
iff the set $K_p(B_0):= \{K_p(x);\quad x\in B_0\}$ is totally bounded (or has compact
closure). Here, $I$ is a set of indexes and $B_0$ is the unit ball in $\RR^n$.
\end{definition}

 We reformulate Proposition 1 of \cite{ammlakmos} here in our context. Its proof depends on
the work of Vogelius and Volkov \cite{volvol}.
\begin{prop} \label{prop1} Let $\l$ be a resonance value of (\ref{res}). Assume that $\gamma_{D_i}$, $i=1,\cdots,m$ be given as in Section 2. Let $T_\epsilon(\omega)$ be defined by (\ref{Teps}) and
$T(\omega)$ by (\ref{T}). Then there exists some neighborhood
$B_\l$ around $\l$ such that for
$\omega\in B_\l$ we have the following
\begin{description}

\item[(a)]  $T_\eps(\omega)$ converges to $T(\omega)$ pointwise.
\item[(b)] For each fixed $\omega$, there exists $\eps_0$ such that the
 set  $\{ T_\eps(\omega) - T(\omega);\ \  \eps <\eps_0 \}$
is collectively compact.
\item[(c)] For each fixed $\omega\in B_\l\setminus \{\l\}$ there exists a
constant $C$ that is independent of
$\eps$ such that for any $f \in H^{1/2}(\partial \Omega)$,
$T_\eps(\omega)^{-1}$ exists and
$$\| T_\eps(\omega)^{-1} f \|_{H^{1/2}(\partial \Omega)} \leq C_\omega
\| f \|_{H^{1/2}(\partial \Omega)}.$$

\item[(d)] The following asymptotic formula holds
\begin{eqnarray}
(T(\omega) - T_\eps(\omega))(f)(x) &=&
\gamma_1 S^\omega(N^\omega-N_\eps^\omega)(f)(x)\nonumber \\
 &=& \ds -\eps^n
\sum_{i=1}^m\gamma_1(1-\frac{\gamma_1}{tr(\gamma_{D_i})}) \n v(z_i)\cdot M^{(i)}
\n_y G(x,z_i) +o(\eps^n)\nonumber \\
\label{fprop1}\end{eqnarray} where $v$ is the solution to
\begin{equation}
\n\cdot\gamma \n v+\omega^2 v=0\ \ \ \ \mbox{in}\ \ \ \ \O \label{vequ}
\end{equation}
$$v=f\ \ \ \ \ \mbox{on} \ \ \ \ \p\O;$$
and the asymptotic term
$o(\eps^n)$ (and its derivatives) are
 uniform in $x\in \p\O$ and $\omega\in B_\l$.
\end{description}
\end{prop}
One also gets the following estimate on pointwise convergence of the
operators and their dual:
\begin{cor} Let $\l$ be a resonance value of (\ref{res}).
$T_\epsilon(\omega)$ be defined by (\ref{Teps}) and
$T(\omega)$ by (\ref{T}). Then there exists some neighborhood
$B_\l$ around $\l$ such that for $f\in H^{1/2}(\p\O)$, $g\in H^{-1/2}(\p\O)$,
and
$\omega\in B_\l$ we have
\begin{equation}
\| (T_\eps(\omega)-T(\omega))f\|_{H^{1/2}(\p\O)}\leq C\eps^n \label{ptws}
\end{equation}
and
\begin{equation}
\| (T^*_\eps(\omega)-T^*(\omega))g\|_{H^{-1/2}(\p\O)}\leq C\eps^n
 \label{dptws}
\end{equation}
where $C$ is independent of  $\eps$ and  uniform for $\omega$ in $B_\l$.
\end{cor}
{\it Proof }\ \ \ \  The first estimate follows from {\bf (d)} of Proposition \ref{prop1}. For the difference of the dual operators,
$$T_\eps^*(\omega)-T^*(\omega)= \gamma_1(N_\eps^{\overline{\omega}}-
N^{\overline{\omega}})S^{-\overline{\omega}}$$
by the same proof as in \cite{ammlakmos} we have a similar asymptotic
formula as above, from which the pointwise convergence estimate follows.
\begin{flushright} $\Box$ \end{flushright}

\section{Asymptotic formulae for the resonances}

Let $\l$ be a pole of $T$, i.e. $T(\l)^{-1}$ does not exist. It is
known, then \cite{kat}, \cite{reesim} that for each
integer $k$, the null space of $N(T(\l)^k)$ is finite dimensional,
and that for $\delta$ small enough, $T^{-1}(z)$ is an analytic
operator in $B_\delta(\l)\setminus\{\l\}$. Let $\alpha$ be the
ascent of $T$, i.e. $\alpha$ is the smallest integer such that
$$N(T(\l)^\alpha)=N(T(\l)^{\alpha+1}).$$ By \cite{kat},
\cite{reesim} such a smallest integer exists and is bigger than or
equal to one. It is the order of $\lambda$ as a pole of
$T^{-1}(z)$ in $B_\delta$. Then the algebraic multiplicity $p$ of
$\l$ is defined by $$p=\mbox{dim} N(T(\l)^\alpha).$$ The geometric
multiplicity $$m = \mbox{dim} N(T)$$  does not clearly not exceed
the algebraic multiplicity $p$, and since $$N(T(\lambda)^{\alpha
-1}) \neq N(T(\lambda)^{\alpha}),$$ both multiplicities  coincide
if and only if the order $\alpha$ of $\lambda$ is equal to one
\cite{how}. It is also known that for any $\delta$ small enough,
there exists $\eps_0$ such that for any $\eps < \eps_0$, $T_\eps$
has exactly $m$ (resp. $p$) resonances  $\{ \l^j_\eps \}$, counted
according to geometric (resp. algebraic) multiplicity, in $B_\delta(\l)$
and such that $T^{-1}_\eps(z)$ exists on $\Gamma=\p B_\delta
(\l)$. Define
\begin{equation}
E= {1\over{2\pi i}}\int_\Gamma T^{-1}(z)dz,\label{E}
\end{equation}
as the projection of $H^{1/2}(\partial \Omega)$ onto the generalized
``eigenspace'' $N(T(\l)^\alpha)$ along $R(T(\lambda)^\alpha)$.
Similarly, define $$E_\eps = {1\over{2\pi i}}\int_\Gamma
T_\eps^{-1}(z)dz$$ which is the projection onto the direct sum of
the generalized eigenspaces of $\l_\eps^j$.  We know
\cite{kat} that for $\eps$ small enough, $$p=\mbox{dim}
N(T(\lambda_\varepsilon)^\alpha) $$ and that for any $u\in R(E) =
N(T(\lambda)^\alpha) $, there exists, for each $\eps$, a $u_\eps
\in R(E_\eps) = N(T(\lambda_\varepsilon)^\alpha)$ such that $$\|
u_\eps -u\|_{H^{1/2}(\partial \Omega)}\rightarrow 0,$$
 i.e., $$\delta
(R(E),R(E_\eps))\rightarrow 0.$$ where $\delta$ is the distance
between the unit balls of the two subspaces. Also, since
$T^{-1}(z)$ is a finitely meromorphic operator in $B_\delta(\l)$
(for $\delta$ small enough), by for example \cite{pros}  we know
that $T^{-1}(z)$ has the expansion
\begin{equation} T^{-1}(z) = {L_\alpha\over{(z-\l)^\alpha}} +
{L_{\alpha-1}\over{(z-\l)^{\alpha-1}}} +\ldots + L_0(z)\label{expT}
\end{equation}
where for $k=1,\dots,\alpha$ each $L_k$ is defined by  $$ L_k = -
T^{k - 1} E$$ and $L_0(z)$ is some analytic operator in ${\cal
L}(H^{1/2}(\partial \Omega))$.

We search now for $\l_\eps$, a pole
of $T_\eps^{-1}(z)$ in $B_\delta (\l)$. To this end, we recall that we know such a pole exists. Then
there exists and $L^2(\p\O)$-normalized $u_\eps\in N(T_\eps)$ such that
$$T_\eps(\l_\eps)u_\eps=0,$$ and hence \be \label{r0}
T(\l_\eps)u_\eps + (T_\eps (\l_\eps)-T(\l_\eps))u_\eps=0. \ee
Assuming that $\l_\eps\neq\l$, $T(\l_\eps)^{-1}$ exists and \be
\label{r1} u_\eps +
T(\l_\eps)^{-1}(T_\eps(\l_\eps)-T(\l_\eps))u_\eps=0. \ee Taking
the $L^2$-inner product with $u_\eps$, we obtain \be \label{r2}
1+\langle T(\l_\eps)^{-1}(T_\eps(\l_\eps)-T(\l_\eps))u_\eps,u_\eps
\rangle=0. \ee $\l_\eps$ is a resonance of the transmission
problem in $B_\delta(\lambda)$ if and only if there exists
$$u_\eps \in H^{1/2}(\partial \Omega)$$ with
$$ \langle u_\eps, u_\eps
\rangle = || u_\eps||^2_{L^2(\partial \Omega)} = 1$$
such that $\l_\eps$ is a root of the complex meromorphic
function \begin{equation}\label{function-g}g(z)=1+\langle
T(z)^{-1}(T_\eps(z)-T(z))u_\eps,u_\eps\rangle.\end{equation}
From now on we concentrate for simplicity  on resonances that are
simple, i.e. with $\alpha = 1$. Note that in this case $m$ does
not necessarily equal one, but $p=m$ and $R(E) = N(T)$. The
derivation of an asymptotic formula in the general case follows
from similar arguments and will be presented at the end of this
paper. We will need the following lemma:
\begin{lem} \label{ll} Assume $\alpha = 1$. Let
 $\{u^j\}_{j=1}^m$ be an $L^2$-orthonormal basis of eigenfunctions
 for $E=N(T(\lambda))$, and let $\{ u^{j*}\}_{j=1}^m$ be the dual basis
for $E^*=N(T^*(\l))$, that is, with
    $$<u^j,u^{i*}>=\delta_{ij}.$$
Then let $u^j_\eps\in N( T_\eps(\l^j_\eps))$ be chosen such that
$$\| u^j_\eps -u^j\|_{H^{1/2}(\partial \Omega)} \rightarrow 0.$$
Then for $z=\l_\eps^j$,
\begin{eqnarray}
\langle L_1(T -T_\eps)
u^j_\eps,u^j_\eps\rangle = \langle (T_\eps-T) u^j,u^{j*}\rangle \nonumber
\\ + o(\| (T-T_\eps)|_{R(E)}\| ) +
o(\| (T-T_\eps)^*|_{R(E^*)}\| ).
\end{eqnarray}
\end{lem}
{\it Proof }\ \ \ \ \  First note that
\begin{eqnarray} \langle L_1(T-T_\eps)u^j_\eps,u^j_\eps\rangle &=&
 \langle (T-T_\eps)u^j_\eps,L_1^*u^j_\eps\rangle\nonumber \\
&=& \sum_i  \langle (T-T_\eps)u^j_\eps,u^{i*}\rangle
\langle u^j_\eps,u^{i*}\rangle. \nonumber
\end{eqnarray}
By adding and subtracting $u^j$ to
$u^j_\eps$, we obtain
\begin{eqnarray} \langle L_1(T-T_\eps)u^j_\eps,u^j_\eps\rangle &=&
\sum_i \langle (T_\varepsilon -T)(u^j_\eps - u^j), u^{i*}\rangle
\langle u^{i*}, u_\varepsilon^j\rangle \nonumber
\\ &+& \sum_i \langle (T_\eps -T)u^j, u^{i*} \rangle \langle
u^{i*},u_\varepsilon^{j} \rangle \nonumber \\ &=& \sum_i \langle
(u^j_\eps - u^j), (T_\varepsilon -T)^* u^{i*}\rangle \langle u^{i*},
u_\varepsilon^j\rangle \nonumber
\\ &+& \sum_i \langle (T_\eps -T)u^j, u^{i*} \rangle \langle
u^{i*},u_\varepsilon^{j} \rangle \nonumber \\
\end{eqnarray}
which, by the orthogonality of the basis functions and the
convergence of $u^j_\eps$, proves the lemma.
\square\\

Using the above  lemma, we may prove the following result.
\begin{prop}\label{ff}
 Let $T$ and $T_\eps$ be the operators defined
by (\ref{T}) and (\ref{Teps}) respectively. Let $\lambda$ be a
resonance of $T$ of order one and multiplicity $m$. For $\delta$
and $\eps$ small enough, $T_\eps$ has exactly $m$ resonances
$(\lambda_\varepsilon^j)$, counted according to multiplicity in
$B_\delta(\lambda)$ and the following asymptotic formula holds:
\begin{eqnarray}
\frac{1}{m} \sum_{j=1}^{m} \l^j_\eps - \l &=& \frac{1}{m}
\sum_{j=1}^m \langle (T_\eps(\l^j_\eps)-T(\l^j_\eps))
u^j,u^{j*}\rangle \nonumber \\&+&
 o(\|(T-T_\eps)|_{R(E)}\| )+
o(\| (T-T_\eps)^*|_{R(E^*)}\| ),\nonumber
\end{eqnarray}
where $E$ is the projection operator defined by (\ref{E}), $E^*$ its
dual, and
$\{ u^j,u^{j*}\}_{j=1}^{m}$ are mutually $L^2$-orthonormal bases
of eigenfunctions of $T(\l)$ and $T^*(\l)$.
\end{prop}

{\it Proof} Let the function $g$ be defined by (\ref{function-g}). Using the
expansion (\ref{expT}), we may rewrite $g(z)$ as
\begin{eqnarray} g(z) &=& 1+ {\langle
L_\alpha(T_\eps(z)-T(z))u_\eps,u_\eps\rangle
\over{(z-\l)^\alpha}}  \nonumber \\ &+&   {\langle L_{\alpha-1}
(T_\eps(z)-T(z))u_\eps,u_\eps\rangle
\over{(z-\l)^{\alpha-1}}} \nonumber \\
&+& \ldots + \langle L_0(z)(T_\eps(z)-T(z))u_\eps,u_\eps\rangle. \nonumber
\end{eqnarray}
Since
$$(\l_\eps-\l)^{\alpha}g(\l_\eps)=0,$$
this gives us
\begin{eqnarray} (\l_\eps-\l)^\alpha &=&
\langle L_\alpha (T(\l_\eps)-T_\eps(\l_\eps))u_\eps,u_\eps\rangle \nonumber \\
&+& (\l_\eps-\l)\langle
L_{\alpha-1}(T(\l_\eps)-T_\eps(\l_\eps))u_\eps,u_\eps\rangle \nonumber\\
&+&\ldots + (\l_\eps-\l)^\alpha \langle L_0(\l_\eps)(T(\l_\eps)-
T_\eps(\l_\eps)u_\eps,u_\eps\rangle.
\nonumber \end{eqnarray}
Bringing the last term over to the left hand side,
\begin{eqnarray} (\l_\eps-\l)^\alpha\left[ 1-\langle L_0
(T-T_\eps)u_\eps,
u_\eps\rangle \right] &=& \langle
L_\alpha (T-T_\eps)u_\eps,u_\eps\rangle \label{formu1}
\\ &+& \ldots + (\l_\eps-\l)^{\alpha-1} \langle
L_1 (T-T_\eps)u_\eps,u_\eps\rangle
\nonumber
\end{eqnarray}
where all operators are evaluated at $z=\l_\eps$. Now define the analytic
function $$h_\eps(z)=\langle L_0(z)(T(z)-T_\eps(z))u_\eps,u_\eps\rangle.$$
We know, from the pointwise convergence of $T_\eps$ to $T$ and the
$H^{1/2}$ convergence of $u_\eps$ to $u$,
that $h_\eps(z)\rightarrow 0$ in $B_\delta$, so
\begin{equation} {1\over{1-h_\eps(z)}}= 1+h_\eps(z) +h_\eps^2(z)+\ldots
\label{formu2} \end{equation} where the expansion is uniform in
$B_\delta$ for $\eps$ small enough. Using Lemma \ref{ll}, (\ref{formu2}), (\ref{formu1}), and averaging
over $j$, the proof holds.
\square\\

The following Lemma describes the dual eigenspace.
\begin{lem}\label{dual} For $j=1, \ldots m$, we may choose
$$u^{j*} = c_{jk}(S^{-\overline{\lambda}})^{-1}(\overline{u^k})$$
where the coefficients are given by
$$c_{jk}=(A^{-1})_{jk}$$
for
$$A_{ki}=\langle (S^{-\overline{\lambda}})^{-1}(\overline{u^k}), u^i
\rangle .$$
\end{lem}
{\it Proof }\ \ \ \ \  First we show that for any $j=1,\cdots,m$,
 $$(S^{-\overline{\lambda}})^{-1}(\overline{u^j})\in N(T^*(\l)).$$
Observe that
$$S^{- \overline{\lambda}} T^*(S^{-
\overline{\lambda}})^{-1}(\overline{u^j}) = (1 -
\frac{\gamma_2}{2}) \overline{u^j} - \gamma_2 S^{-
\overline{\lambda}} (D^\lambda)^*(S^{-
\overline{\lambda}})^{-1}(\overline{u^j}) + \gamma_1 S^{-
\overline{\lambda}} N^{\overline{\lambda}} (\overline{u^j}). $$
From $T(\lambda)(u^j) =0$ it follows that
$$S^{-
\overline{\lambda}} T^*(S^{-
\overline{\lambda}})^{-1}(\overline{u^j}) = \gamma_2 ( D^{-
\overline{\lambda}} - S^{- \overline{\lambda}} (D^\lambda)^*(S^{-
\overline{\lambda}})^{-1})(\overline{u^j}).$$ The Calder\'on 's
commutation relation $S D = D^* S$ yields  $$S^{-
\overline{\lambda}} T^*(S^{-
\overline{\lambda}})^{-1}(\overline{u^j}) = 0,$$ and hence
$$T^*(S^{- \overline{\lambda}})^{-1}(\overline{u^j}) = 0.$$
In addition, for
$$u^{j*} = c_{jk}(S^{-\overline{\lambda}})^{-1}(\overline{u^k}),$$
we have the desired orthogonality properties
$$\langle u^{j*}, u^i\rangle =\delta_{ij}$$
\square\\

On the other hand, we may compute the first term on the right hand side
of Proposition \ref{ff} to prove the following main result.

\begin{thm}\label{thm-main1} Assume that we have (\ref{zass}), and all hypothesis of Proposition \ref{ff}. Let $\gamma_{D_i}$ be given as in Section $2$. Then, the following asymptotic formula holds:
\begin{eqnarray}
\frac{1}{m} \sum_{j=1}^{m} \l^j_\eps - \l &=&
 -\eps^n \frac{1}{m}\sum_{i=1}^{m}\gamma_1(1-\frac{\gamma_1}{tr(\gamma_{D_i})})\sum_{j,l=1}^m \n u^j(0)\cdot M^{(i)}
\n c_{jl}u^l(0)
+o(\eps^n)\nonumber
\end{eqnarray}
where the coefficients are given by
$$c_{jl}=(A^{-1})_{jl}$$
for
$$A_{li}=\langle (S^{-\overline{\lambda}})^{-1}(\overline{u^l}), u^i
\rangle .$$
\end{thm}
{\it Proof} Let us now compute the first term on the right hand side
of Proposition \ref{ff}. Consider one term in the summation
corresponding to a particular $j$. For clarity of exposition we
will temporarily neglect the $j$ superscript on $\l_\eps$, $u$,
and $u^*$. We have
\begin{eqnarray} \langle (T_\eps (\l_\eps)-T(\l_\eps))u,u^{*} \rangle
&=& \langle\gamma_1
S^{\l_\eps}(N_\eps^{\l_\eps}-N^{\l_\eps})u,u^{*} \rangle
\nonumber
\\ &=& \langle\gamma_1
(N_\eps^{\l_\eps}-N^{\l_\eps})u,S^{-\overline{\l}_\eps}
u^{*} \rangle \nonumber
\\ &=&\int_{\p\O}
\gamma_1(N_\eps^{\l_\eps}-N^{\l_\eps})u(x)
\overline{S^{-\overline{\l}_\eps}u^{*}}(x)d\sigma_x. \nonumber
\end{eqnarray}
Recall that
$$N^{\l_\eps}u
={\p\alpha^\eps\over{\p\nu}}$$ and
$$N_\eps^{\l_\eps}u={\p\beta^\eps\over{\p\nu}}$$ where
$\alpha_\eps$ solves
\begin{eqnarray}
\gamma_1\Delta\alpha^\eps +(\l_\eps)^2\alpha^\eps &=& 0\ \ \ \
\mbox{in} \ \ \ \ \O \nonumber \\ \alpha^\eps &=& u \ \ \ \ \
\mbox{on}\ \ \ \ \p \O \nonumber
\end{eqnarray}
and $\beta^\eps$ solves
\begin{eqnarray}
\n\cdot\gamma_\eps\n\beta^\eps +(\l_\eps)^2\beta^\eps &=& 0\ \ \
\ \mbox{in} \ \ \ \ \O \nonumber \\ \beta^\eps &=& u \ \ \ \ \
\mbox{on}\ \ \ \ \p \O. \nonumber
\end{eqnarray}
From  (\ref{fprop1}), we have the uniform asymptotic expansion for
$y$ on $\p\O$,
\be
\nonumber
{\p\beta^\eps\over{\p\nu}}-{\p\alpha^\eps\over{\p\nu}}=-
\eps^n\sum_{i=1}^{m}(1-\frac{\gamma_1}{tr(\gamma_{D_i})})\n \alpha^\eps(0)\cdot M^{(i)}
 {\p\over{\p\nu_y}}\n_x{\hat N}^{(\l_\eps)}(0,y) +o(\eps^n).
 \ee
where have now noted the
dependence of ${\hat N}$ on $\omega$ explicitly. Since
$\lambda_\eps \rightarrow \lambda$ and $u$ satisfies $\gamma_1
\Delta u + \lambda^2 u = 0$ in $\Omega$ we can show that
 $$|\n\alpha^\eps(0)-\n u(0)|\rightarrow
0. $$
Therefore we obtain
\be
\label{r3}
{\p\beta^\eps\over{\p\nu}}-{\p\alpha^\eps\over{\p\nu}}=
-\eps^n\sum_{i=1}^{m}(1-\frac{\gamma_1}{tr(\gamma_{D_i})})\n u(0)\cdot M^{(i)}
 {\p\over{\p\nu_y}}\n_x{\hat N}^{(\l_\eps)}(0,y) +o(\eps^n).
 \ee
Define $w_\eps\in H^1(\O)$  by
 $$\gamma_1\Delta w_\eps +\overline{\l} w_\eps=0 \ \ \ \ \mbox{in}\ \ \ \O$$
 $$w_\eps=S^{-\overline{\l_\eps}}u^{*}\ \ \ \mbox{on}\ \ \p\O.$$
Then by using (\ref{r3}) we deduce that
\begin{eqnarray}
\label{r4} \langle (N_\eps^{\l_\eps}-N^{\l_\eps})u, w_\eps\rangle
 &=&
\langle{\p\beta^\eps\over{\p\nu}}-{\p\alpha^\eps\over{\p\nu}},
w_\eps\rangle \nonumber \\ &=& -\eps^n \sum_{i=1}^{m}(1-\frac{\gamma_1}{tr(\gamma_{D_i})}) \n u(0)\cdot M^{(i)} \n \overline{w}(0) \nonumber
\\ & &
+o(\eps^n) \nonumber
\end{eqnarray}
where $w$ is the $H^1$ limit of $w_\eps$, that is, (returning to
using the $j$ superscript), $w^j$ satisfies $$\gamma_1\Delta
w^j+\overline{\l} w^j=0 \ \ \ \ \mbox{in}\ \ \ \O$$
$$w^j=S^{-\overline{\l}}u^{j*}\ \ \ \mbox{on}\ \ \ \ \p\O.$$ Lemma
\ref{dual} gives that
$$\overline{w}^j=c_{jk}u^k,$$
from which we obtain
the next theorem.
\square\\

In the case where the resonance $\lambda$ is not simple ($\alpha >1$),
we can easily generalize this result to the following theorem.
\begin{thm} \label{thg}
Let $T$ and $T_\eps$ be the operators defined
by (\ref{T}) and (\ref{Teps}) respectively. Let $\lambda$ be a
resonance of $T$ of order (ascent) $\alpha$, geometric multiplicity $m$
and algebraic multiplicity $p$. Let  $\{u^j\}$ be an $L^2(\p\O)$-orthonormal
basis for $N(T)$.
Then for $\delta$ and $\eps$ small enough, $T_\eps$ has exactly $m$
resonances $\{ \lambda_\varepsilon^j\}$, counted according to
geometric multiplicity in $B_\delta(\lambda)$. For  $j=1 \ldots
m$  the following asymptotic formula holds:
\begin{eqnarray}
(\l^j_\eps - \l)^\alpha &=&-
 \eps^n \sum_{i,l=1}^{m}\gamma_1(1-\frac{\gamma_1}{tr(\gamma_{D_i})}) \n u^j(0)\cdot M^{(i)} \n
c_{jl}u^l(0) \nonumber \\  &+&
o(\eps^n),\nonumber
\end{eqnarray}
where the coefficients are given by
$$c_{jl}=(A^{-1})_{jl}$$
for
$$A_{li}=\langle (S^{-\overline{\lambda}})^{-1}(\overline{u^l}), u^i
\rangle .$$
 \end{thm}
{\it Proof }
From the same proof as Lemma \ref{ll},
we can show that for any $k=1,\ldots, \alpha$,
 \begin{eqnarray} \langle L_k(T-T_\eps) u^j_\eps,u^j_\eps\rangle &=&
 \langle (T-T_\eps)u^j,u^{j*}\rangle \nonumber \\
&+& o(\| T-T_\eps|_{R(E)}\| ) +
 o(\| (T-T_\eps)^*|_{R(E^*)}\| ).\nonumber
\end{eqnarray}
For $k=1,\ldots,\alpha-1$,  it also follows that
\begin{eqnarray}
 (\l^j_\eps-\l)^{\alpha-k}\langle L_k(T-T_\eps) u^j_\eps,u^j_\eps\rangle
\nonumber &=&
 o(\| T-T_\eps|_{R(E)}\| ) +
 o(\| (T-T_\eps)^*|_{R(E^*)}\| ).\nonumber
\end{eqnarray}
Using this fact with (\ref{formu2}) and (\ref{formu1}) to obtain that
$$ (\l^j_\eps - \l)^\alpha=
  \langle (T-T_\eps)u^j,u^{j*}\rangle +o(\eps^n)$$
which achieves the proof of the theorem.
\square

\end{document}